\documentclass[letterpaper]{article}
\pdfoutput=1
\usepackage{aaai}
\usepackage{times}
\usepackage{helvet}
\usepackage{courier}

\usepackage{multirow}
\usepackage{subfigure}
\usepackage{verbatim}

\usepackage[pdftex]{graphicx}
\usepackage[pdftex]{color}
\usepackage{url}

\def\etal{et~al.\,}

\nocopyright

\title{Network Analysis of Recurring YouTube Spam Campaigns}
\author{Derek O'Callaghan, Martin Harrigan, Joe Carthy, P\'{a}draig Cunningham\\
School of Computer Science \& Informatics, University College Dublin\\
\{derek.ocallaghan,martin.harrigan,joe.carthy,padraig.cunningham\}@ucd.ie}
\begin{document}
\maketitle
\begin{abstract}

As the popularity of content sharing websites such as YouTube and Flickr has
increased, they have become targets for spam, phishing and the distribution of
malware. On YouTube, the facility for users to post comments can be used by
spam campaigns to direct unsuspecting users to bogus e-commerce websites. In
this paper, we demonstrate how such campaigns can be tracked over time using
network motif profiling, i.e. by tracking counts of indicative network motifs. By
considering all motifs of up to five nodes, we identify discriminating motifs
that reveal two distinctly different spam campaign strategies. One of these
strategies uses a small number of spam user accounts to comment on a large
number of videos, whereas a larger number of accounts is used with the other. 
We present an evaluation that uses motif profiling to track two active campaigns 
matching these strategies, and identify some of the associated user accounts.

\end{abstract}

\section{Introduction}

The usage and popularity of content sharing websites continues to rise each
year. For example, the number of Flickr uploads has risen to a total of six
billion images, having increased annually by 20\% over the past five
years\footnote{\url{http://news.softpedia.com/news/Flickr-Boasts-6-Billion-Photo-Uploads-215380.shtml}}.
Similarly, YouTube now receives more than three billion views per day, with
forty-eight hours of video being uploaded every minute; increases of 50\%
and 100\% respectively over the previous 
year\footnote{\url{http://youtube-global.blogspot.com/2011/05/thanks-youtube-community-for-two-big.html}}.
Unfortunately, such increases have also resulted in these sites becoming more
lucrative targets for spammers hoping to attract unsuspecting users to malicious
websites, where a variety of threats such as scams (phishing, e-commerce) and
malware can be found. This is a particular problem for YouTube given its
facility to host discussions in the form of video comments~\cite{DBLP:journals/corr/abs-1103-5044}.
Opportunities exist for the abuse of this feature with the availability of
bots\footnote{\url{http://youtubebot.com/}} that can be used to post spam
comments in large volumes.

\begin{figure}
	\begin{center}
		\mbox{
			\subfigure{\includegraphics{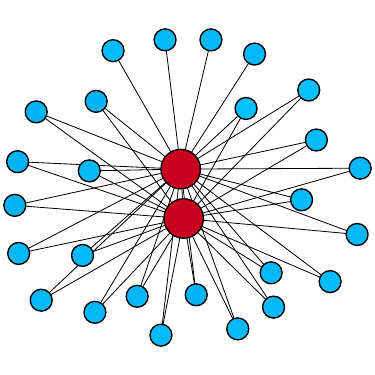}} \qquad
			\subfigure{\includegraphics{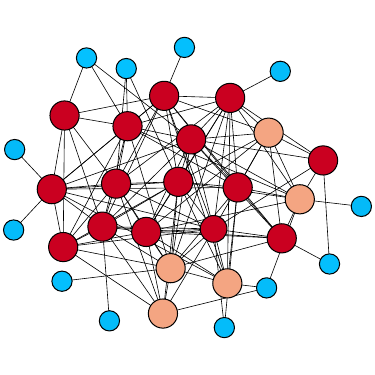}}
			}
	\end{center}
	\caption{Strategies of two spam campaigns targeting YouTube in 2011 - small
	number of accounts each commenting on many videos (left), and larger number of accounts
	each commenting on few videos (right). Blue nodes are videos, red nodes are
	accounts marked as spam, beige nodes are spam accounts not marked accordingly.}
	\label{fig:campaigns}
\end{figure}

Our investigation has found that bot-posted spam comments are often associated
with orchestrated campaigns that can remain active for long periods of time,
where the primary targets are popular videos. Such campaigns tend to employ a
variety of detection evasion techniques, such as variants of the same
fundamental message content, perhaps with different website domains, and an
ever-evolving set of fake user accounts. An initial manual analysis of data
gathered from YouTube revealed activity from a number of campaigns, two of which
can be seen in Figure~\ref{fig:campaigns}. The results presented in this paper
confirm the presence of these campaigns, along with their recurring nature.

As an alternative to traditional
approaches that attempt to detect spam on an individual (comment) level (e.g.
domain blacklists), this paper presents an evaluation of the detection of these
recurring campaigns using network analysis, based on networks derived from the
comments posted by users to videos. This approach uses the concept of
\textit{network motif profiling}~\cite{Milo25102002,Milo05032004,Wu:2011:CWP:2065023.2065036}, where motif counts
from the derived networks are tracked over time. Given that different campaign
strategies can exist (see Figure~\ref{fig:campaigns}), the objective is to
discover certain discriminating motifs that can be used to identify particular
strategies and the associated users as they periodically recur.

This paper begins with a description of related work in the domain. The
collection of contemporary YouTube data, comprised of comments posted to the
most popular videos over a period of time, is then discussed. Next, the
methodology used by the detection approach is described in detail, from
derivation of the comment-based networks to the subsequent network motif profile
generation. The results of an experiment for a seventy-two hour period are then
presented. These results demonstrate the use of certain discriminating motifs to
identify some of the users associated with two separate campaigns we have
discovered within this time period. Further analysis of the campaign websites is also
provided. Finally, the overall conclusions are discussed, and some suggestions
for future work are made.

\section{Related Work}

\subsection{Structural and spam analysis}

The network structure of YouTube has been analysed in a number of separate
studies. Paolillo~\etal~\shortcite{Paolillo08} investigated the social structure
with the generation of a user network based on the friendship relationship,
focusing on the degree distribution. They found that YouTube is similar to other
online social networks with respect to degree distribution, and that a social
core exists between authors (uploaders) of videos. An alternative network based
on related videos was analysed by Cheng~\etal~\shortcite{Cheng08:4539688}. Given
that the resulting networks were not strongly connected, attention was reserved
for the largest strongly connected components. These components were found to
exhibit \textit{small-world} characteristics~\cite{small-world-watts-strogatz},
with large clustering coefficients and short characteristic path lengths,
indicating the presence of dense clusters of related videos.

Benevenuto~\etal~\shortcite{Benevenuto:2008:UVI:1459359.1459480} created a
directed network based on videos and their associated responses. Similarly, they
found that using the largest strongly connected components was more desirable
due to the large clustering coefficients involved. This was a precursor to
subsequent work concerned with the detection of \textit{spammers} and
content \textit{promoters} within YouTube~\cite{Benevenuto:2008:IVS:1451983.1451996,Benevenuto:2009:DSC:1571941.1572047}.
Features from the video responses networks (e.g. clustering coefficient,
reciprocity) were used as part of a larger set to classify users
accordingly. Other YouTube spam investigations include the recent work of Sureka~\shortcite{DBLP:journals/corr/abs-1103-5044}, based on the detection of
spam within comments posted to videos. A number of features were derived to
analyse the overall activity of users, rather than focusing on individual
comment detection.

An extensive body of work has been dedicated to the analysis of spam within
other online social networking sites. For example, Mishne~\etal~\shortcite{Mishne05blockingblog} suggested an approach for the detection of link
spam within blog comments using the comparison of language models. Gao~\etal~\shortcite{Gao:2010:DCS:1879141.1879147} investigated the proliferation of spam
within Facebook ``wall" messages, with the detection of spam clusters using
networks based on message similarity. This particular study demonstrated the
\textit{bursty} (recurring) and \textit{distributed} aspects of
botnet-driven spam campaigns, as discussed by Xie~\etal~\shortcite{Xie:2008:SBS:1402958.1402979}. The shortcomings of URL blacklists
for the prevention of spam on Twitter were highighted by Grier~\etal~\shortcite{Grier:2010:SUC:1866307.1866311}, where it was found that
blacklist update delays of up to twenty days can occur. This is a particular
problem with the use of shortened URLs, the nature of which was recently
analysed by Chhabra~\etal~\shortcite{Chhabra:2011:PPL:2030376.2030387}.

\subsection{Network motif analysis}

\textit{Network motifs}~\cite{Milo25102002,ShenOrr02} are structural patterns
in the form of interconnected $n$-node subgraphs that are considered to be
inherent in many varieties of network, such as biological, technological and
sociological networks. They are often used for the comparison of said networks,
and can also indicate certain network characteristics. In particular, the work of
Milo~\etal~\shortcite{Milo05032004} proposed the use of \textit{significance}
profiles based on the motif counts found within networks to enable the
comparison of local structure between networks of different sizes. In this case,
the generation of an ensemble of random networks was required for each
significance profile. An alternative to this
approach~\cite{Wu:2011:CWP:2065023.2065036} involved the
use of motif profiles that did not entail random network generation. Instead,
profiles were created on an \textit{egocentric} basis for the purpose of
characterising individual \textit{egos}, encompassing the motif counts from the
entirety of egocentric networks within a particular network.

The domain of spam detection has also profited from the use of network motifs or
subgraphs. Within a network built from email addresses~\cite{Boykin:2005:1432647}, a low clustering coefficient (based on the number of
triangle structures within a network) may indicate the presence of spam
addresses, with regular addresses generally forming close-knit communities, i.e.
a relatively higher number of triangles. Becchetti~\etal~\shortcite{Becchetti:2008:ESA:1401890.1401898} made use of the number of
triangles and clustering coefficient as features in the detection of web spam.
These two features were found to rank highly within an overall feature set.
Motifs of size three (triads) have also been used to detect spam comments in
networks generated from blog interaction~\cite{Kamaliha:2008:CNM:1490299.1490781}. It was found that certain motifs were
likely to indicate the presence of spam, based on comparison with corresponding
random network ensembles.

Separately, network motifs have also been used to characterize network traffic~\cite{Allan:2009:UNM:1811982.1812090}. A 
network was created for each application (e.g. HTTP, P2P applications), and
nodes within the network were classified using corresponding motif profiles.


\section{Data Collection}

Following the lead of earlier related YouTube research, a data set was
collected in order to permit the investigation of contemporary spam comment
activity. An extensive crawl of the YouTube network was performed by other
researchers~\cite{Paolillo08,Benevenuto:2009:DSC:1571941.1572047}. In our case,
we opted for a specific selection of the available data given that spam comments
in YouTube tend to be directed towards a subset of the entire video set, i.e. more
popular videos generally have a higher probability of attracting attention from
spammers, thus ensuring a larger audience. This characteristic has also been
seen on other online social networks such as Twitter~\cite{twitter-spam-benevenuto}.

Another issue to be considered is the accessibility of certain YouTube data
attributes. The \textit{recent activity} of a user profile contains a number of
potential attributes for use in the derivation of representative networks, such
as comments posted to videos, and subscriptions added to other users.
Similarly, the list of subscribers for a particular user would also be
useful. However, access to these attributes can often be restricted, meaning
that reliance on such data may lead to inaccuracies during subsequent
experiments. On the contrary, comments (and the users who posted them) found on
a public video's page are always accessible. Given these issues, we decided 
to use only data to which access was not restricted, namely the
comments posted to videos along with the associated user accounts.

\subsection{Retrieval process}

The data has been retrieved using the YouTube Data
API\footnote{\url{http://code.google.com/apis/youtube/getting_started.html#data_api}}.
This API provides access to video and user profile information. There are some limits
associated with using the API, of which further details are provided below.
Apart from video and user information, access is also provided to standard feeds
such as \textit{Most Viewed} videos, \textit{Top Rated} videos etc. The fact
that these feeds are periodically updated (usually daily) facilitates our
objective of analysing recurring spam campaigns, as it enables the retrieval of
popular videos (i.e. those attracting spam comments) on a continual basis.
Therefore, the retrieval process is executed periodically as follows:

\begin{enumerate}
    \item Retrieve the current video list from the \textit{most viewed} standard
    feed for the US region (the API limits this to a maximum of 100 videos).
    \item For each video in the list:
    \begin{enumerate}
        \item If this video has not appeared in an earlier feed list, retrieve
        its meta-data such as upload time, description etc.
        \item Retrieve the comments and associated meta-data for the last
        twenty-four hours, or those posted since the last retrieval time (if
        more recent). The API limits the returned comments to a maximum of 1,000.
    \end{enumerate}
    \item In order to track the comment activity on particular videos appearing
    intermittently in the \textit{most viewed} feed, comments are also
    retrieved for those videos not in the current feed list that appeared in an earlier
    list from the previous forty-eight hours.
\end{enumerate}

\subsection{Data set properties}

Data retrieval began on October 31st, 2011, and details of the videos
and comments as of January 17th, 2012 can be found in
Table~\ref{tab:datasetprops}\footnote{The data set is available at
\url{http://mlg.ucd.ie/yt}
}.

\begin{table}[h]
\begin{center}
\begin{tabular}{| l | r |}
\hline
Videos & 6,407 \\ \hline
Total comments & 6,431,471 \\ \hline
Comments marked as spam & 481,334 \\ \hline
Total users & 2,860,264 \\ \hline
Spam comment users & 177,542 \\ \hline 
\end{tabular}
\end{center}
\caption{Data set properties}
\label{tab:datasetprops}
\end{table}

An interesting feature of the API is the \textit{spam hint} property provided
within the video comment meta-data. This is set to \textit{true} if a comment
has previously been marked as spam, either by the spam filter or manually with the
``Flag for spam'' button available with each comment posted on a video's page.
However, this property cannot be considered reliable due to its
occasional inaccuracy, where innocent comments can be marked as spam, while
obvious spam comments are not marked as such. This will be demonstrated later
in the results discussion. Similar evidence of this property's unreliability was
also encountered in earlier work~\cite{DBLP:journals/corr/abs-1103-5044}.

Although the comment spam hint is used for approximate annotation of the data
(Table~\ref{tab:datasetprops}), it is not relied upon as a label for the purposes of
this evaluation. Other research in this area~\cite{Benevenuto:2009:DSC:1571941.1572047} performed manual label annotation of
YouTube data for use in subsequent classification experiments. An accurate
annotation process will be considered in future work.

\section{Methodology}

\subsection{Comment processing and network generation}

Our methodology requires the generation of a network to represent the comment
posting activity of users to a set of videos. Initially, comments made during a
specified time interval are selected from the data set discussed in the
previous section. However, a number of pre-processing steps must be executed
before an appropriate network can be generated similar to those in Figure~\ref{fig:campaigns}.

Spammers try to obfuscate the text of comments from a
particular campaign in order to bypass their detection by any filters.
Obfuscation techniques include the use of varying amounts of additional
characters (e.g. whitespace, Unicode newlines, etc.) within the comment text, or
different textual formations (e.g. additional words, misspellings) of the same fundamental
message. Some examples of these can be seen in the next section. 

To counteract these efforts, each comment is converted to a set of tokens.
During this process, stopwords are removed, along with any non-Latin-based words
as the focus of this evaluation is English-language spam comments.
Punctuation characters are also removed, and letters are converted to lowercase.
A modified comment text is then generated from the concatenation of the
generated tokens. As initial analysis found that spam comments can often be
longer than regular comments, any texts shorter than a minimum length (currently
25 characters) are removed at this point. Although the campaign strategies under
discussion here are concerned with attracting users to remote sites through the
inclusion of URLs in comment text, comments without URLs are currently retained.
This ensures the option of analysing other types of spam campaigns, such as
those encouraging channel views, i.e. \textit{promoters}~\cite{Benevenuto:2009:DSC:1571941.1572047}, along with the behaviour of regular
users.

A network can then be generated from the remaining modified comment texts. This
network consists of two categories of node, \textit{users} and \textit{videos}.
An undirected edge is created between a user and a video if at least one comment
has been posted by the user on the video, where the edge weight represents the
number of comments in question. For the moment, the weight is merely recorded
but is not subsequently used when counting motifs within the network. To capture
the relationship between the users involved in a particular spam campaign,
undirected and unweighted edges are created between user nodes based on the
similarity of their associated comments. Each modified (tokenized) comment
text is converted to a set of hashes using the Rabin-Karp rolling hash method~\cite{Karp:1987:ERP:1012156.1012171}, with a sliding window length of 3. A
pairwise distance matrix, based on Jaccard distance, can then be generated from
these comment hash sets. For each pairwise comment distance below a threshold
(currently 0.6), an edge is created between the corresponding users if one does
not already exist. 

Afterwards, any users whose set of adjacent nodes consists
solely of a single video node are removed. Since these users have commented on
only one video, and are in all likelihood not related to any other users, they are
not considered to be part of any spam campaign. The resulting network tends to
consist of one or more large connected components, with a number of
considerably smaller connected components based on videos with
a relatively minor amount of comment activity. Finally, an approximate
labelling of the user nodes is performed, where users are labelled as spam users if they
posted at least one comment whose \textit{spam hint} property is set to true. All
remaining users are labelled as regular users. Although this can lead to label
inaccuracies, the results in the next section demonstrate that such inaccuracies
will be perceivable.

\subsection{Network motif profiles}

Once the network has been generated, a set of \textit{egocentric} networks can
be extracted. In this context, given that the focus is on user activity, an
\textit{ego} is a user node, where its egocentric network is the induced 
\textit{k-neighbourhood} network consisting of those user and video nodes whose
distance from the ego is at most \textit{k} (currently 2). Motifs from size three to five
within the egocentric networks are then enumerated using FANMOD~\cite{Wernicke01052006}. A set of motif counts is maintained for each ego, where
a count is incremented for each motif instance found by FANMOD that contains the
ego.

A network motif count profile is then created for each ego. As the number of
possible motifs can be relatively large (particularly if directed and/or
weighted edges are considered), the length of this profile will vary for each
network generated from a selection of comment data, rather than relying upon a
profile with a (large) fixed length. For a particular generated network, the
profiles will contain an entry for each of the unique motifs found in the
entirety of its constituent egocentric networks. Any motifs not found for a
particular ego will have a corresponding value of zero in the associated motif
profile.

As mentioned previously, the work of Milo~\etal~\shortcite{Milo05032004}
proposed the generation of a \textit{significance} profile, where the
significance of a particular motif was calculated based on its count in a
network along with that generated by an ensemble of corresponding random
networks. These profiles then permitted the subsequent comparison of different
networks. In this work, the egocentric networks are compared with each other,
and the generation of random ensembles is not performed. An alternative
\textit{ratio} profile $rp$~\cite{Wu:2011:CWP:2065023.2065036} is created for
each ego, where the ratio value for a particular motif is based on the counts
from all of the egocentric networks, i.e.:

\begin{equation}
	rp_i = \frac{nmp_i - \overline{nmp_i}}{nmp_i + \overline{nmp_i} + \epsilon}
\end{equation}

\medskip
Here, $nmp_i$ is the count of the $i^{th}$ motif in the ego's motif profile,
$\overline{nmp_i}$ is the average count of this motif for all motif profiles,
and $\epsilon$ is a small integer that ensures that the ratio is not
misleadingly large when the motif occurs in only a few egocentric networks. To
adjust for scaling, a normalized ratio profile $nrp$ is then created for each
ratio profile $rp$ with:

\begin{equation}
	nrp_i = \left(\frac{rp_i}{\sum{rp_i^2}}\right)^{\frac{1}{2}}
\end{equation}

\medskip
The generated set of normalized ratio profiles usually contain correlations
between the motifs. Principal components analysis (PCA) is used to adjust for
these, acting as a dimensionality reduction technique in the process.
We can visualize the first two principal components as a starting
point for our analysis. This is discussed in the next section.

\section{Experiments and Results}

\begin{figure*}
	\begin{center}
		\mbox{
			\subfigure{\includegraphics{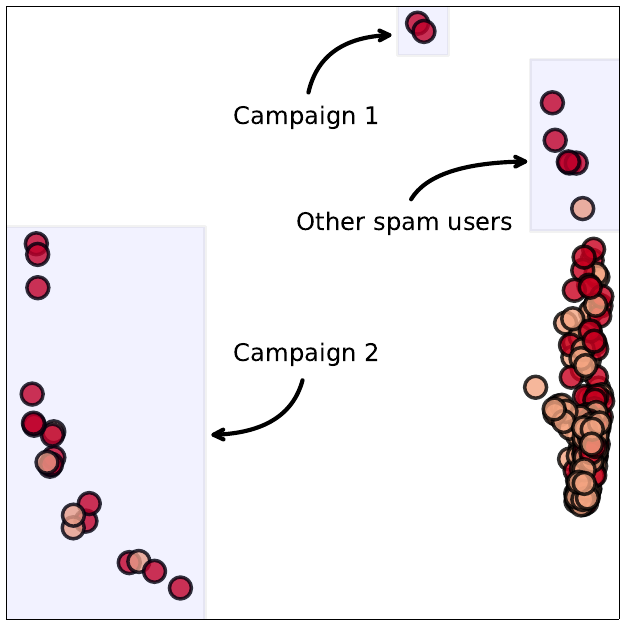}}
			\qquad \qquad
			\subfigure{\includegraphics{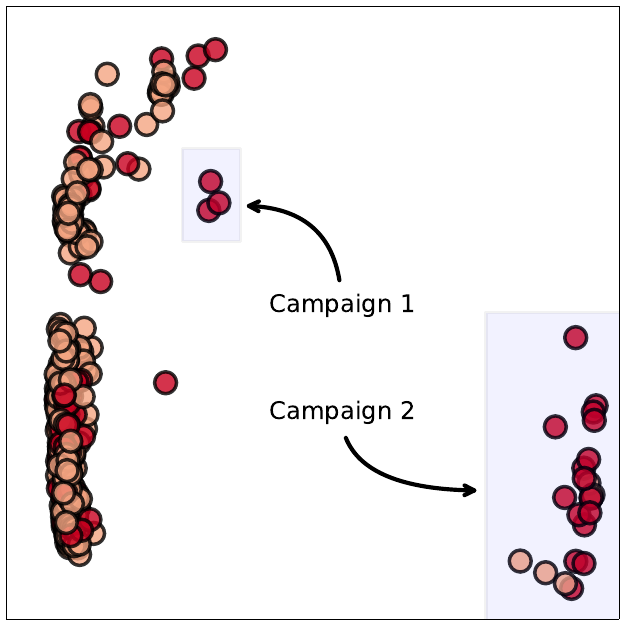}}
			}
	\end{center}
	\caption{Spatialization of the first two principal components of the normalized
	ratio profiles for Windows 10 and 11 (red nodes are users with comments marked as spam, beige
	nodes are all other users). Both spam campaigns are highlighted.}
	\label{fig:pca}
\end{figure*}

For the purpose of this evaluation, the experiments were focused upon tracking
two particular spam campaigns that we discovered within the data set. The
campaign strategies can be seen in Figure~\ref{fig:campaigns}, i.e.
a small number of accounts each commenting on many videos (Campaign~1), and a
larger number of accounts each commenting on few videos (Campaign~2). A period
of seventy-two hours was chosen where these campaigns were active, starting on
 November 14th, 2011 and ending on November 17th, 2011.

In order to track the campaign activity over time, this period was split into
twelve windows of six hours each. For each of these windows, a network of user
and video nodes was derived using the process described in the previous section.
A normalized ratio profile was generated for each ego (user), based on the motif
counts of the corresponding egocentric network. Principal components analysis
was then performed on these profiles to produce 2-dimensional spatializations of
the user nodes, using the first two components. These spatializations act
as the starting point for the analysis of activity within a set of time windows.

\subsection{Visualization and initial analysis}

Having inspected all twelve six-hour windows, two windows containing
activity from both campaigns have been selected for detailed analysis
here. These are from November 17th, 2011; Window~10 (04:19:32 to
10:19:32) and Window~11 (10:19:32 to 16:19:32). Their derived network details
can be found in Table~\ref{tab:windows}.

\begin{table}[h]
\begin{center}
\begin{tabular}{| r | r | r | r |}
\hline
Window & Video nodes & User nodes (spam) & Edges \\ \hline\hline
10 & 263 & 295 (107) & 907 \\ \hline
11 & 296 & 523 (137) & 1627 \\ \hline
\end{tabular}
\end{center}
\caption{Network details for Windows 10 and 11}
\label{tab:windows}
\end{table}

A spatialization of the first two principal components of the normalized ratio
profiles for these windows can be found in Figure~\ref{fig:pca}. Users posting
at least one comment marked as spam (using the \textit{spam hint} property) are
in red, all other users are in beige. The points corresponding to the spam
campaign users have been highlighted accordingly. From the spatializations, it
can be seen that in both windows:

\begin{enumerate}
    \item The vast majority of users appear as overlapping points in larger
    clusters (on the right and left respectively).
    \item There is a clear distinction between the two different campaign
    strategies, as these points are plotted separately (both from regular users
    and each other). 
    \item The inaccuracy of the \textit{spam hint} comment property is
    demonstrated as the Campaign~2 clusters contain users not coloured in red,
    i.e. none of their comments were marked as spam (further details of
    these users can be seen in Figure \ref{fig:motifusers}). Similarly, the
    reverse is true with the larger clusters of regular users.
\end{enumerate}

Apart from the highlighted campaign clusters, other spam nodes in the
spatializations have been correctly marked as such. For example, the five users
that are separated from the normal cluster in Window~10 (``Other spam
users") appear to be isolated spam accounts having similar behaviour to the
Campaign~1 strategy, but on a smaller scale. This also applies to the single
separated user in Window~11. They are not considered further during this
evaluation as they are not part of a larger campaign.

Further analysis of Campaign~1 revealed that two and three users
were active in Windows 10 and 11 respectively (five separate users), posting the
following comments:

\begin{center}
\begin{tabular}{l }

{\small \textsf{Three most cool things in the World for me before}}\\
{\small \textsf{1 ))))) Jordan--the super star}}\\
{\small \textsf{2 ))))) 66cheap. com--the cheapest shopping site}}\\
{\small \textsf{3 ))))) the iphone -- best connector}}\\
{\small \textsf{NOW THERE'S ONE MORE, IT'S THE VIDEO}}\\
{\small \textsf{ABOVE!!!!!!!!!!}}
\\   
\\
{\small \textsf{Three Best things in the World for me now: ): ): ): ): ) }}\\
{\small \textsf{1. Lily------My boyfriend！}}\\
{\small \textsf{2. 55cheap. com--the cheapest shopping site}}\\
{\small \textsf{3. the video above---- the most ironical and interesting}}\\
{\small \textsf{video I think:]:]:]:]:] :]:] :]:]}}

\end{tabular}
\end{center}

\begin{figure*}
	\begin{center}$
		\begin{array}{cc}
			\includegraphics{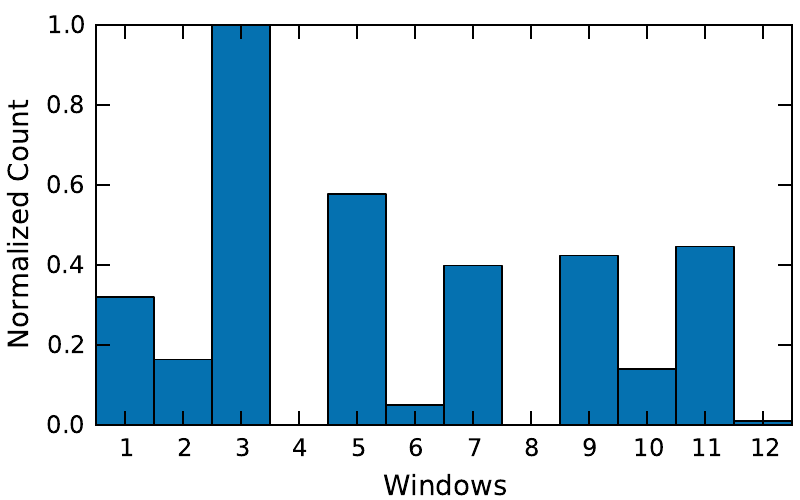}
			&
			\includegraphics{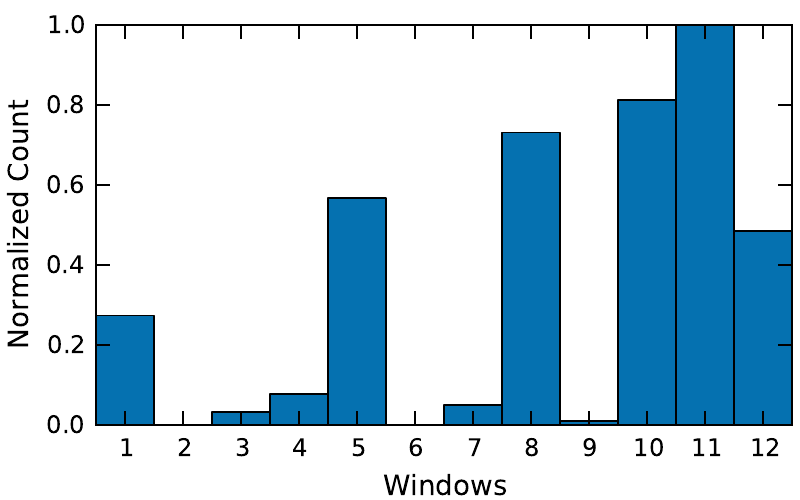}
			\\
			\qquad \quad
			\includegraphics{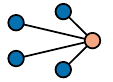} &
			\qquad \quad
			\includegraphics{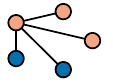}
			\end{array}$
	\end{center}
	\caption{Tracking the recurring activity of Campaign~1 (left) and Campaign
	2 (right) for all six-hour windows from 14th November 2011 to 17th November
	2011, using a single discriminating motif for each campaign.}
	\label{fig:motifwindows}
\end{figure*}

\medskip
Although there are certain differences between these comments, they are clearly
from the same campaign. This behaviour is also seen in both windows with
Campaign~2, featuring a larger number of users, although there are fewer occurrences of
identical comments. Nevertheless, a similarity is noticeable, for example:

\medskip
\begin{center}
\begin{tabular}{ l }

{\small \textsf{Don't miss this guys, the CEO of apple is releasing }}\\
{\small \textsf{ipads on Thursday: osapple.co.nr}}
\\   
\\
{\small \textsf{dont miss out. November 17 - new apple ceo is shipping}}\\
{\small \textsf{out old ipad and iphones}}\\
{\small \textsf{Not a lie. Go to this webpage to see what I mean:}}\\
{\small \textsf{bit.ly\textbackslash vatABm}}
\\ 
\end{tabular}
\end{center}

\medskip
Both of these comments are made by the same user in different windows. However,
while the first comment was accurately marked as spam, the second was
\textbf{not}. An assumption here could be that the URL in the first comment is
on a spam blacklist, while the shortened URL in the second enables such a list
to be bypassed. Similar shortcomings are discussed in earlier work~\cite{Chhabra:2011:PPL:2030376.2030387}.

\subsection{Discriminating motifs}

An inspection of the individual motif counts found that certain
motifs have relatively higher counts for users involved in the spam campaigns,
than those found for regular users. These motifs may be considered indicative
of different campaign strategies, and a subset can be found in Table~\ref{tab:motifs}.

\begin{table}[h]
\begin{center}
\begin{tabular}{| c | c | c |}
\hline
 & Campaign~1 & Campaign~2 \\ \hline
\multirow{12}{*}{Motifs} 
&  &  \\
&
\includegraphics{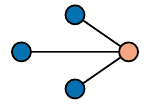}
&
\includegraphics{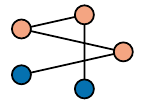}
\\
&  &  \\
&
\includegraphics{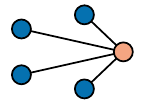}
&
\includegraphics{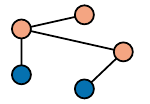}
\\
&  &  \\
&  &
\includegraphics{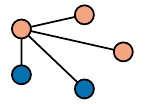}
\\ 
&  &  \\
\hline
\end{tabular}

\end{center}
\caption{A subset of discriminating motifs for different spam campaign
strategies (user nodes are beige, video nodes are blue).}
\label{tab:motifs}
\end{table}

These discriminating motifs would appear to correlate with the existing
knowledge of the campaign strategies. Campaign~1 consists of a small number of
users commenting on a large number of videos, and so it would be expected
that motifs containing only one user node with a large number of video
node neighbours have higher counts for the users involved, as is the case here.
The motifs considered indicative of Campaign~2 are more subtle, in that the
number of user and video nodes is similar, with both user and video nodes
present in the set of neighbours for a particular user. However, all three
highlight the fact that users appear to be more likely to be connected to other
users rather than videos. This makes sense given that with this campaign, a
larger number of users tend to comment on a small number of videos each, and
the potential for connectivity between users is higher given the similarity of
their comments. These motifs would also appear to indicate that users in the
campaign don't comment on the same videos, as no two users share a video node
neighbour.

Figure~\ref{fig:motifwindows} contains plots for the counts of two of these
motifs for each of the six-hour windows. The counts were normalised using the
edge count for the corresponding window networks followed by min-max
normalization. The fluctuation in counts across the windows appears to track the
recurring periodic activity of these campaigns, as confirmed by separate analysis of the
data set. This would appear to corroborate the \textit{bursty} nature of spam
campaigns~\cite{Xie:2008:SBS:1402958.1402979,Gao:2010:DCS:1879141.1879147}.

\begin{figure*}
	\begin{center}$
		\begin{array}{cc}
			\includegraphics{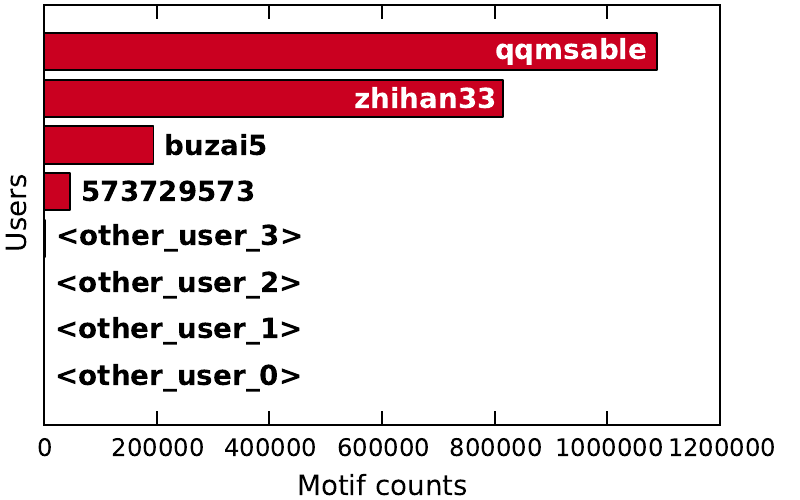}
			&
			\includegraphics{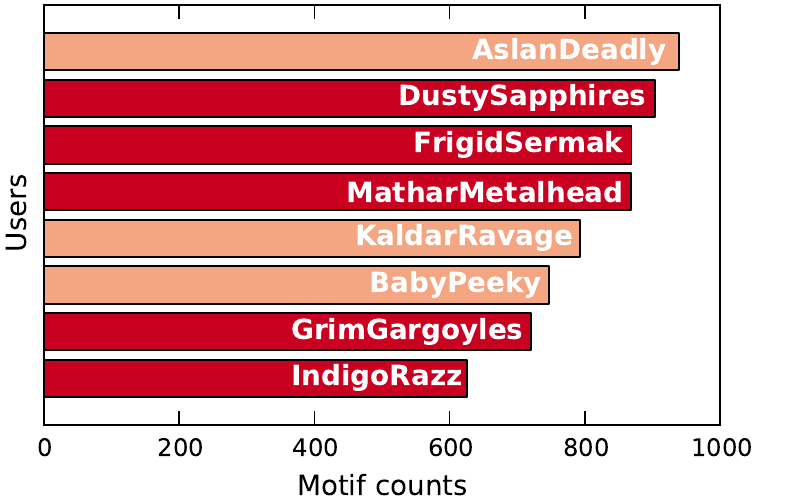}
			\\
			\includegraphics{results/20111207_171634/5144-1user_4video_4edges_u0v0_u0v1_u0v2_u0v3_small.pdf} &
			\includegraphics{results/20111207_171634/5324-3user_2video_4edges_u0u2_u1u2_u2v0_u2v1_small.pdf}
			\end{array}$
	\end{center}
	\caption{Users associated with Campaign~1 (left) and Campaign~2
	(right), having the highest counts for a single discriminating
	motif for each campaign from Window~11 (17th November 2011 10:19:32 to
	16:19:32). Note how three of the users in Campaign~2 are coloured differently,
	i.e. none of their comments were marked as spam. Users not involved in the
	campaigns have been anonymized.}
	\label{fig:motifusers}
\end{figure*}

Finally, Figure~\ref{fig:motifusers} plots the user counts in descending
order for these two motifs in Window~11. With the Campaign~1 motif (left), the
first four users are involved and have considerably higher counts than the
remaining users. There are also differences in counts between the campaign
users themselves, indicating the most active users in this window. All users
plotted for the Campaign~2 motif (right) are indeed involved. Three of the
Campaign~2 users were coloured differently to the others, highlighting the fact
that none of their comments were accurately marked as spam. These same three
users can be seen in the right spatialization in Figure \ref{fig:pca}.

\section{Campaign Analysis}

Following the inspection of the discriminating motifs, the websites and domains
associated with the comments posted by the campaign 1 users were then analyzed.
The following domains were found in the data set in comments beginning with
\textit{``Three Best things''} and \textit{``Three most cool things''} (as seen
in the example comments listed in the previous section), and can be categorized
as follows:

\begin{enumerate}
    \item National Football League (NFL) merchandise: \textbf{2006jerseys.com},
    \textbf{66cheap.com}, \textbf{shopofnfl.com}.
    \item Footwear: \textbf{21boots.com}.
    \item Wider range of merchandise (e.g. clothing, accessories):
    \textbf{36shopping.com}, \textbf{55cheap.com}, \textbf{55goods.com}.
\end{enumerate}

It is quite clear that all of these sites
are related given the high similarity between them, e.g. various index page
titles containing the text \textit{``The Cheapest Shopping Site''},  identical
payment options and the same contact email address. There are also some
inconsistencies in the HTML content, for example, some 66cheap.com pages refer
to shopofnfl.com and jerseysofnfl.com, and 55cheap.com pages refer to
36shopping.com. Suspicious claims are also made, such as \textit{``SHOPOFNFL.COM
was the online shop of NFL''}. At first glance, 21boots.com looks different to
the others, but further investigation reveals similarities such as the payment
options. The domains appear to have been registered by the same
person\footnote{\url{http://whois.domaintools.com}}. As 55cheap.com has been
previously identified as a known scam
website\footnote{\url{http://answers.yahoo.com/question/index?qid=20110426143143AArdrbK}},
it is safe to assume that all of these sites should be treated as such.

Further analysis of 55goods.com shows it to be an older site, as its About page
alleges that it has been in operation for \textit{``17 years''} since 1993. This
would suggest that this scam has been in operation since 2010 at the very least. It
appears that the About page details contain further inconsistencies, e.g.,
55goods.com states that \textit{``In 2009, 78.8\% of our annual revenue was from
the international market\ldots''}, while 55cheap.com, allegedly in operation for
\textit{``18 years''} since 1993 contains the same statement with merely a
change in year: \textit{``In 2010, 78,8\% of our annual revenue was from the
international market\ldots''}.

A total of 24 different user accounts were used to send the associated comments
found in the data set. Although some of these accounts have been suspended by
YouTube, others remain active. The campaign appears to rotate the 
existing accounts for comment posting, and new accounts are created
on a continual basis. The four accounts for this campaign listed in Figure~\ref{fig:motifusers} are currently active as of January 2012. Of these four, the
oldest account was created in August 2011, while the most recent was created in
October 2011.

\section{Conclusions and Future Work}

YouTube spam campaigns typically involve a number of spam bot user accounts
controlled by a single spammer targeting popular videos with similar comments
over time. We have shown that dynamic network analysis methods are effective for
identifying the recurring nature of different spam campaign strategies, along
with the associated user accounts. We have used a characterization of YouTube
users in terms of motifs in the comment network to highlight the users in
question. While the YouTube comment scenario could be characterized as a network
in a number of ways, we use a network representation comprising user and video
nodes, user-video edges representing comments and user-user edges representing
comment similarity.

The discriminating power of these motif-based characterizations can be seen in
the PCA-based spatialization in Figure~\ref{fig:pca}. It is also clear from
Figure~\ref{fig:motifwindows} that histograms of certain discriminating motifs
show the level of activity in the two campaign strategies over time.
Furthermore, counts of these motifs in the egocentric networks of users
highlight the associated accounts (Figure~\ref{fig:motifusers}).

\subsection{Future Work}

For future experiments, it will be necessary to annotate the data set with
spam/non-spam labels, or perhaps a more extensive annotation that considers the
associated campaign strategies. Feature selection of a subset of motifs could
then be performed along with subsequent user classification. The use of a subset
of motifs is attractive, as it would remove the current requirement to count all
motif instances found in the user egocentric networks, which can be a lengthy
process.

\section{Acknowledgements}
This work is supported by 2Centre, the EU funded Cybercrime Centres of
 Excellence Network and  Science Foundation Ireland under grant 08/SRC/I140: Clique: Graph and Network Analysis Cluster.

\bibliography{main}
\bibliographystyle{aaai}
\end{document}